\documentclass[runningheads]{rnl}
\usepackage[latin1]{inputenc}
\usepackage[francais]{babel}
\usepackage{epsfig}

\begin{document}

\title*{Henri Poincaré et l'émergence du~concept de cycle limite}


\titretab{Henri Poincaré et l'émergence du~concept de cycle limite \protect\newline}


\titrecourt{Henri Poincaré et l'émergence du~concept de cycle limite}

\author{Ginoux Jean-Marc}

\index{Ginoux Jean-Marc}              

\auteurcourt{J.--M. Ginoux}

\adresse{Laboratoire PROTEE, I.U.T. of Toulon, University of South,\\
B.P. 20132, 83957, La Garde Cedex, France.}

\email{ginoux@univ-tln.fr, http://ginoux.univ-tln.fr}

\maketitle

\begin{resume}
Le concept de \og{}cycle limite\fg{} fut introduit par Henri Poincaré dans son second mémoire \og{}Sur les courbes définies par une équation différentielle\fg{} en 1882. Du point de vue de la Physique, un cycle limite stable (ou attractif) correspond à la représentation de la solution périodique d'un système (mécanique ou~électrique) dissipatif dont les oscillations sont entretenues par le système lui-même. Inversement, l'existence d'un cycle limite stable~garantit l'entretien des oscillations. Jusqu'à~présent, l'historiographie considérait que le mathématicien russe Aleksandr' Andronov avait été le tout premier à établir une telle correspondance entre la solution périodique d'un système auto-oscillant et le concept de cycle limite de Poincaré. La découverte récente d'une série de conférences réalisées par Henri Poincaré en 1908 à l'Ecole Supérieure des Postes et Télégraphes (aujourd'hui Telecom Paris Tech)~démontre qu'il avait déjà mis en application son concept de cycle limite pour établir l'existence d'un régime stable d'ondes entretenues dans un dispositif de la~T.S.F.\footnote{Télégraphie Sans Fil.} Cet article a donc pour objet d'une part de retracer l'émergence de ce concept depuis sa création par Poincaré et, d'autre part de mettre en évidence l'importance de son rôle dans l'histoire des oscillations non linéaires.

\end{resume}


\section{Henri Poincaré et le concept de cycle limite}
\label{}

Dans le chapitre VI de son second mémoire \og{}Sur les courbes
définies par une équation différentielle\fg{}, Poincaré \cite[p. 261]{P2} présente la \og{}Théorie des cycles limites\fg{}. En faisant
appel à la \og{}Théorie des conséquents\fg{} qu'il a préalablement exposée et qui contient en essence le principe de ce que l'on appelle aujourd'hui une \og{}section de Poincaré\fg{}, il démontre l'existence d'un nouveau genre de courbes fermées\footnote{Les courbes fermées correspondant aux solutions de type centre ayant été exclues de la discussion par Poincaré.} qu'il nomme
\og{}cycle limite\fg{}. Au chapitre VII, Poincaré \cite[p. 274]{P2}
présente alors le tout premier exemple de cycle limite. Il s'agit du
système de deux équations différentielles du premier ordre et du
premier degré suivant~:

\begin{equation}
\frac{dx}{x\left( {x^2+y^2-1} \right)-y\left( {x^2+y^2+1}\right)}=\frac{dy}{y\left( {x^2+y^2-1} \right)+x\left( {x^2+y^2+1} \right)}
\label{eq1}
\end{equation}

\begin{figure}[htbp]
\centerline{\includegraphics[width=5.61cm,height=6.27cm]{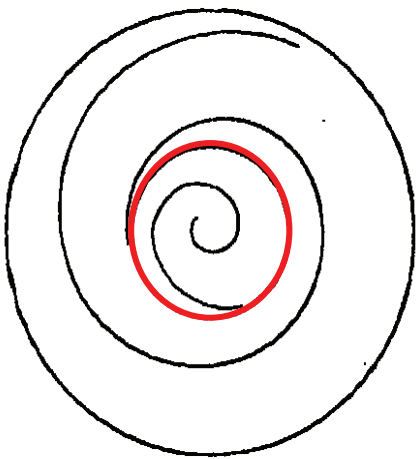}}
\caption[Premier exemple de cycle limite]{Premier exemple de cycle limite, d'après Poincaré \cite[p. 280]{P2}}
\label{fig1}
\end{figure}

Bien entendu ce système qui a été construit \textit{ad hoc} par Poincaré
pour illustrer son propos ne recouvre pas une réalité physique.
Néanmoins, il permet de mettre en évidence l'existence d'une courbe
fermée invariante (au sens de Darboux \cite{Darboux}) qui n'est autre que le
cercle cycle limite algébrique d'équation (voir Fig. \ref{fig1})~:

\[
x^2+y^2=1
\]

Cependant, comme le rappelle aussitôt Poincaré \cite[p. 283]{P2}:

\begin{quote}

\og{}Quand les cycles limites ne sont pas algébriques, une discussion
complète est évidemment impossible~; car on ne pourra jamais trouver
en termes finis l'équation des cycles limites.\fg{}

\end{quote}

Sur la Fig. \ref{fig1}, Poincaré a représenté le cycle limite
d'équation $x^2 + y^2 = 1$ et les courbes trajectoires décrivant un point
mobile \og{}qui ne pourra jamais le franchir et qui restera toujours à
l'intérieur de ce cycle, ou toujours à l'extérieur\fg{}.

Dans la \textit{Notice sur les Travaux scientifiques d'Henri Poincaré} faite par lui-même en 1884, Poincaré \cite[p. 25]{Pnot} fournit une définition
mathématique très précise de ce concept~:

\begin{quote}

\og{}[{\ldots}] il y a un autre genre de courbes fermées qui jouent un
rôle capital dans cette théorie~: ce sont les \textit{cycles limites}. J'appelle ainsi les
courbes fermées qui satisfont à notre équation
différentielle et dont les autres courbes définies par la même
équation se rapprochent asymptotiquement sans jamais les atteindre.
Cette seconde notion n'est pas moins importante que la première.
Supposons, en effet, que l'on ait tracé un cycle limite~; il est clair
que le point mobile dont nous parlions plus haut ne pourra jamais le
franchir et qu'il restera toujours à l'intérieur de ce cycle, ou
toujours à l'extérieur.\fg{}

\end{quote}

\begin{figure}[htbp]
\centerline{\includegraphics[width=8.56cm,height=8.56cm]{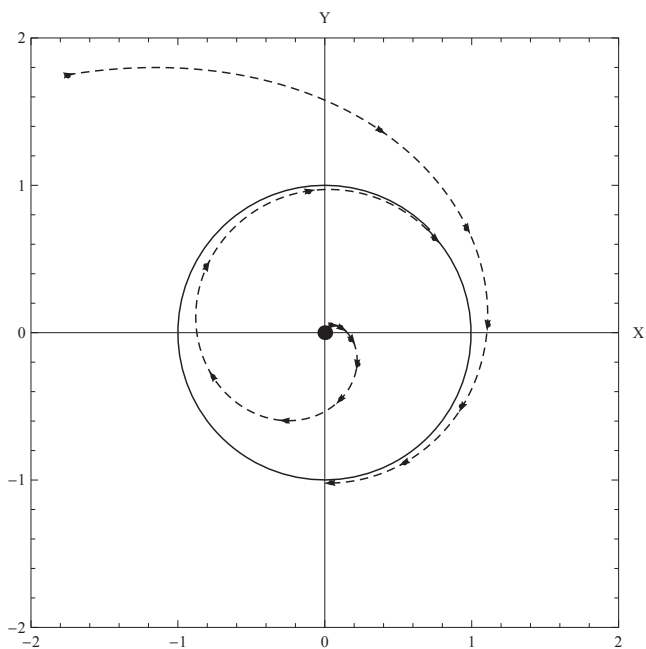}}
\caption{Illustration du concept de cycle limite.}
\label{fig2}
\end{figure}

Poincaré imagine la solution d'une équation différentielle comme
un point mobile (une planète par exemple) décrivant une trajectoire
dans le plan (de phase). Sa définition signifie que la trajectoire
décrite par cette solution prend la forme d'une courbe fermée sur
elle-même (cercle noir sur la Fig. \ref{fig2}) qui attirerait toute autre
trajectoire se trouvant aussi bien à l'intérieur qu'à
l'extérieur (trajectoires en pointillés sur la Fig. \ref{fig2}).

Du point de vue mathématique, si l'on excepte le cas particulier
utilisé par Poincaré comme exemple, les cycles limites constituent
un nouveau type de \og{}courbes définies par une équation
différentielle\fg{} et non par une équation cartésienne comme
celle du cercle.\\

Du point de vue de la physique, cela implique que quelles que soient les
conditions initiales choisies, la trajectoire s'approche asymptotiquement de
la courbe fermée sans jamais l'atteindre. Si cette explication
légitime la dénomination de \textit{cycle limite} elle ne fournit cependant aucune
indication quant à sa signification et il faudra attendre un quart de
siècle pour que Poincaré lui donne sa véritable
interprétation en 1908.\\

Lorsque l'on s'intéresse à l'analyse des phénomènes
oscillatoires on commence généralement par étudier le pendule
dont on représente (modélise) les oscillations au moyen d'une
équation différentielle. On démontre alors que dans le cas du
pendule sans frottements les oscillations sont périodiques mais
dépendent des conditions initiales alors que dans celui du pendule avec
frottements les oscillations sont amorties.

A la fin du XIX\ieme{} siècle, la \textit{Théorie des Oscillations Linéaires} qui décrit ce type
d'oscillations, avait atteint son apogée lorsque l'on découvrit un dispositif électromécanique présentant un curieux comportement
oscillatoire qu'elle n'était pas en mesure d'expliquer.

\section{Jean-Marie Anatole Gérard-Lescuyer et la machine série-dynamo}

En 1880, un ingénieur électricien du nom de Jean-Marie Anatole
Gérard-Lescuyer eut l'idée d'associer une \textit{machine dynamo-électrique} jouant le rôle de
générateur à une \textit{machine magnéto-électrique} pouvant être assimilée dans ce cas
à un moteur. Gérard-Lescuyer \cite[p. 226]{GL} explique ainsi le phénomène qu'il observe :

\begin{quote}

\og{}[{\ldots}]~si l'on envoie le courant produit par une machine
dynamo-électrique dans une machine magnéto-électrique, on
assiste à un phénomène étrange.

Aussitôt que le circuit est fermé, la machine
magnéto-électrique se met en mouvement~; elle tend à prendre une
vitesse de régime, en rapport avec l'intensité du courant qui
l'anime~; mais subitement elle se ralentit, s'arrête et repart en sens
contraire, pour s'arrêter de nouveau et tourner dans le même sens
que précédemment. En un mot, elle est animée d'un mouvement
alternatif régulier, qui dure autant que le courant qui l'actionne.\fg{}

\end{quote}

Constatant donc l'inversion périodique du sens de rotation de la machine
magnéto-électrique, alors que le courant est continu, il s'interroge
sur la cause de ce phénomène oscillatoire \og{}étrange\fg{}. Ne
parvenant pas à trouver une explication, il écrit~:

\begin{quote}

\og{}Que devons-nous conclure ?~Rien, sinon que nous nous trouvons en
présence d'un paradoxe scientifique, dont l'explication se fera, mais
qui ne laisse pas d'être intéressant.\fg{}

\end{quote}

En réalité, le phénomène que Gérard-Lescuyer vient de
mettre en évidence c'est l'existence de frottements non linéaires.
Ces frottements ne sont plus, comme dans le cas du pendule, proportionnels
à l'amplitude des oscillations mais sont représentés par une
fonction non linéaire de l'amplitude (une fonction cubique par exemple).
Ceci implique que lorsque les frottements deviennent négatifs ils
entretiennent les oscillations au lieu de les amortir.

Ce dispositif peut être considéré comme l'un des premiers
exemples de système dissipatif, c'est-à-dire, de système dont
les oscillations sont entretenues par le système lui-même. Plus
tard, on parlera de \textit{systèmes auto-entretenus} et d'\textit{oscillations auto-entretenues} ou d'\textit{auto-oscillations}\footnote{Voir Ginoux \cite{Ginthesis} et Ginoux {\&} Lozi \cite{GinouxLozi}.}. Les oscillations auto-entretenues par de tels systèmes
seront également représentées par une équation
différentielle dont la solution périodique prendra la forme d'un
cycle limite de Poincaré. Une correspondance biunivoque sera alors
établie entre la solution périodique d'un système auto-entretenu
et le concept de cycle limite dont l'existence garantira l'entretien ou la
persistance des oscillations. Persistance en effet, car il arrivait
également à cette époque que des ingénieurs cherchent à
faire cesser les oscillations intempestives qui prenaient naissance au sein
de certains dispositifs mécaniques. Ce fut le cas d'Henry Léauté
qui fut confronté à ce genre de problème en 1885 et qui fit
appel, pour le résoudre, au concept de \textit{cycle limite} mais sans faire cependant aucun
lien avec les travaux de Poincaré \cite{P1,P4}.

\section{Henry Léauté et la régulation des machines hydrauliques}

En 1885, l'ingénieur Henry Léauté (1847-1916) publie un long
mémoire intitulé~: `` Sur les oscillations à longue période
dans les machines actionnées par des moteurs hydrauliques et sur les
moyens de prévenir ces oscillations\fg{} \cite{Leaut}. A
cette époque les ingénieurs en hydraulique sont confrontés à
un délicat problème dans les systèmes de vannage. En effet, le
moteur employé pour l'ouverture ou la fermeture de la vanne produit par
rétroaction des oscillations intempestives engendrant un
phénomène de battement des plus préjudiciables. Ces oscillations
dont la période est de l'ordre de quelques dizaines de secondes sont
alors appelées~: \og{}oscillations à longue période\footnote{ Pour
plus de détails voir la thèse de Remaud \cite[p. 143 et
suivantes]{Remaud}.}\fg{}. Pour résoudre ce problème Léauté va faire
appel au concept de cycle limite élaboré trois ans auparavant par
Henri Poincaré \cite[p. 261]{P2} sans avoir hélas réalisé la
correspondance qu'il venait d'établir entre Science et Technique. Le
mathématicien Camille Jordan (1838-1922) chargé de rédiger sa
nécrologie rappela ainsi ses recherches~:

\begin{quote}

\og{}Les oscillations à longue période, si redoutables dans les
machines hydrauliques, ont également attiré l'attention de M.
Léauté. Elles n'avaient été étudiées avant lui que
pour les régulateurs à action directe. M. Léauté a
traité le cas o\`{u} la régularisation intervient par l'action d'une
vanne. Construisant alors une courbe ayant pour abscisses l'ouverture de la
vanne et pour ordonnée la vitesse correspondante de la machine, il a
reconnu que ces oscillations se produisent seulement lorsque ladite courbe
est fermée. L'intégration de l'équation différentielle du
problème lui fait conna\^{\i}tre les cas o\`{u} cette circonstance-se
présente\footnote{Voir Jordan \cite[p. 501]{Jordan}}.\fg{}

\end{quote}

\begin{figure}[htbp]
\centerline{\includegraphics[width=11.09cm,height=6.38cm]{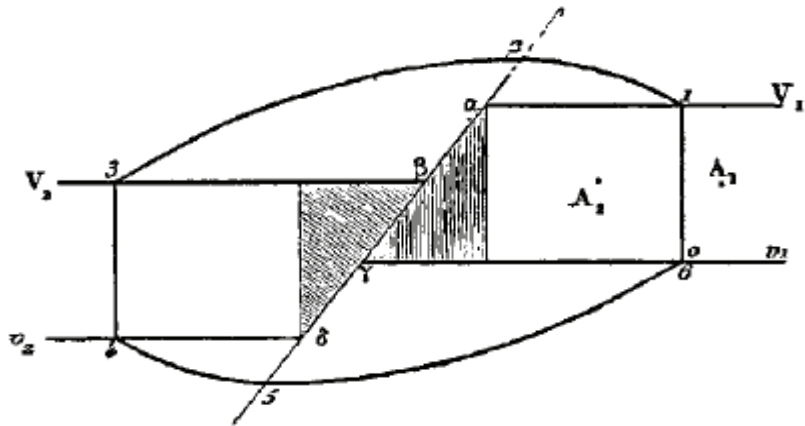}}
\caption[Cycle fermé]{Cycle fermé, d'après Léauté \cite[p.77]{Leaut}}
\label{fig3}
\end{figure}

\begin{figure}[htbp]
\centerline{\includegraphics[width=9.738cm,height=6.453cm]{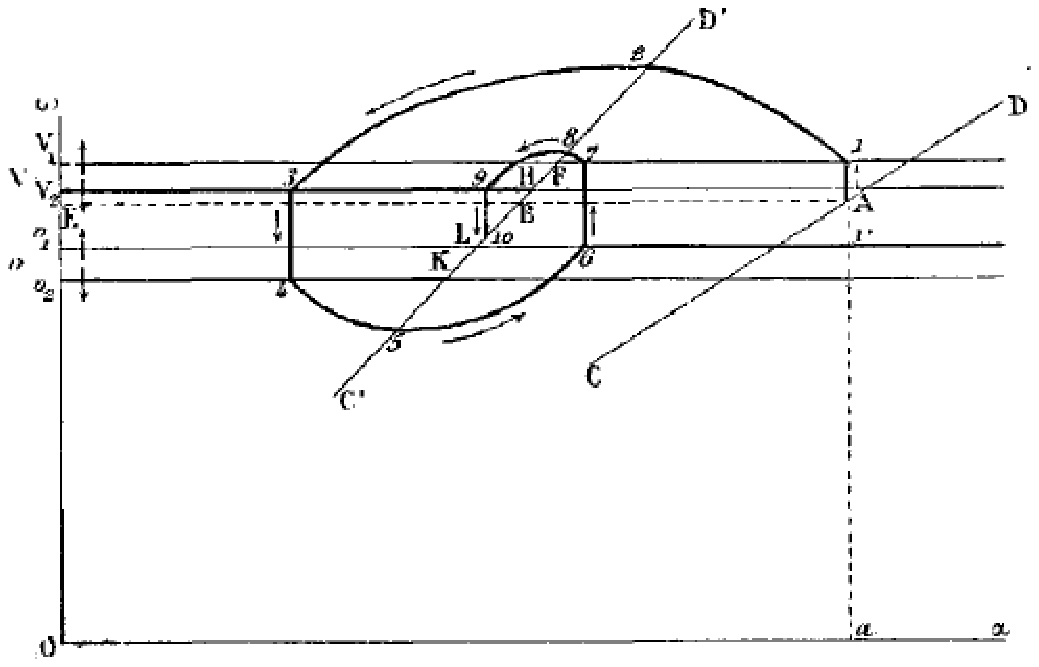}}
\caption[Cycle de Léauté]{Cycle de Léauté \cite[p. 10]{Leaut}}
\label{fig4}
\end{figure}

Pour résoudre ce problème de régulation dans les machines
actionnées par des moteurs hydrauliques, Léauté propose donc de
représenter dans le plan de phase ayant pour abscisse l'ouverture de la
vanne et pour ordonnée la vitesse correspondante de la machine la courbe
figurant l'évolution des oscillations (voir Fig. \ref{fig3} et Fig. \ref{fig4}). Au second
paragraphe du Chapitre IV intitulé \og{}Propriétés du cycle
fermé\fg{}, Léauté décrit les caractéristiques d'un cycle
limite de Poincaré.

En effet, il considère tout d'abord que l'état initial
de la machine peut \^{e}tre représenté par un point extérieur
(A$_{1})$ ou un point intérieur (A$_{2})$. Puis, Léauté \cite[p. 77]{Leaut} explique que

\begin{quote}

\og{}Les cycles successifs\footnote{Ce que Léauté nomme des cycles
successifs correspond à des courbes trajectoires intégrales de
l'équation différentielle caractérisant l'évolution des
oscillations dans ce problème.} décrits à partir de A$_{1}$
étant assujettis à tendre vers le cycle fermé et ayant tous un
effet utile ou un effet nuisible, on voit que ce sera forcément l'effet
utile qui se produira.\fg{}

\end{quote}

Il ajoute alors que~les cycles successifs ne peuvent franchir le cycle
fermé puis étudie de la m\^{e}me manière le cas d'un point à
l'intérieur. Il en conclut :

\begin{quote}

\og{}Ce fait explique les phénomènes qui résultent de l'existence
du cycle fermé et rend compte ainsi de la production des oscillations
à longues périodes. Toute la question est donc ramenée à
fixer les conditions sous-lesquelles il n'y aura pas de cycle fermé.\fg{}

\end{quote}

Bien que Léauté se situe dans une problématique exactement
inverse de tous ceux qui vont lui succéder puisqu'il recherche les
conditions pour lesquelles il n'y a pas de cycle fermé, Léauté
réalise ainsi une correspondance du m\^{e}me type de celle
qu'établira Poincaré \cite{P5,P6} vingt ans plus tard entre solution
périodique d'un oscillateur et cycle fermé, $i.e.$ \textit{cycle limite}.

Trente ans plus tard, les ingénieurs Léon Barbillion\footnote{Pour
une biographie de Barbillion voir la thèse de Remaud \cite{Remaud}.} et Paul
Cayère \cite{Barb} feront appel à la méthode inaugurée par
Léauté \cite{Leaut} pour résoudre le problème de la régulation
(indirecte) des groupes électrogènes actionnés par des turbines
hydrauliques (voir Fig. 5).

\begin{figure}[htbp]
\centerline{\includegraphics[width=14.193cm,height=6.192cm]{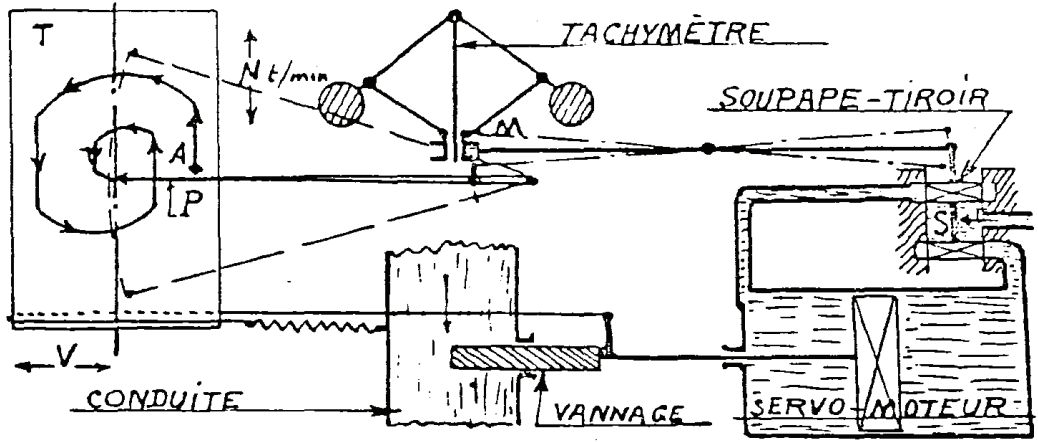}}
\caption[Enregistrement du cycle de Léauté]{Enregistrement du cycle de Léauté, d'après Cayère
[1958, pl. 2457] (voir \cite{Remaud})}
\label{fig5}
\end{figure}

Barbillion et Cayère \cite[p. 747]{Barb} publient leur travail dans la \textit{Revue Générale de l'électricité} entre le 25 mai et le 1$^{er}$ juin 1918
qui débute par cette phrase~:

\begin{quote}

\og{}Lorsqu'on étudie, dans de nombreux travaux publiés sur la
question, l'important problème de la régulation indirecte des
groupes électrogènes actionnés par des turbines hydrauliques, on
constate que les solutions employées actuellement dérivent toutes
d'une fa\c{c}on plus ou moins directe des principes exposés en 1885 par
M. Léauté dans son \og{}Mémoire sur les oscillations à longue
période dans les machines actionnées par des moteurs hydrauliques et
sur les moyens de prévenir ces oscillations\fg{}.\fg{}
\end{quote}

\begin{figure}[htbp]
\centerline{\includegraphics[width=10.45cm,height=10.8cm]{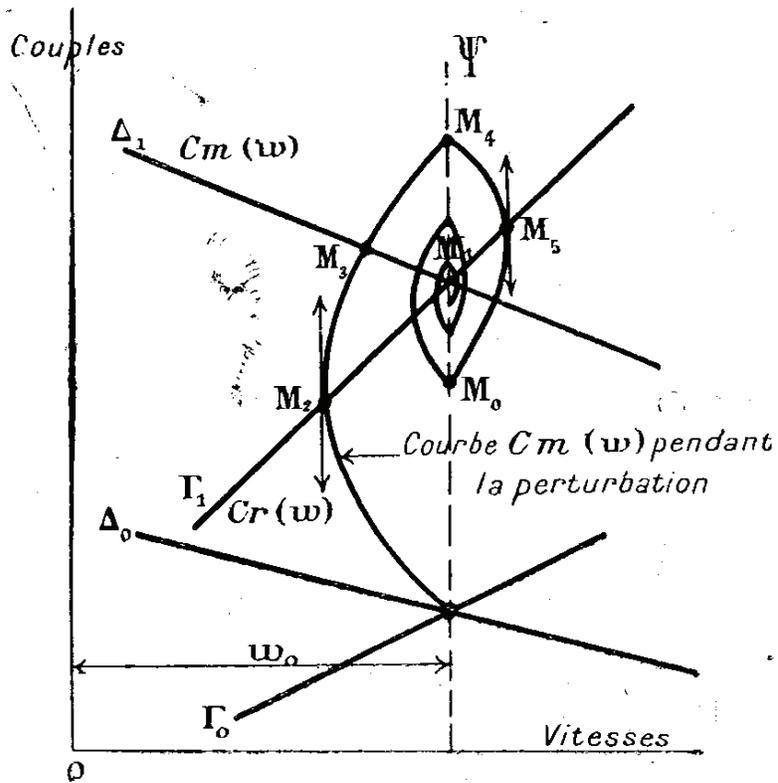}}
\caption[Caractéristique du régulateur indirect asservi]{Caractéristique du régulateur indirect asservi, d'après Barbillion et Cayère \cite[p. 761]{Barb}}
\label{fig6}
\end{figure}

Barbillion et Cayère \cite[p. 761]{Barb} reproduisent sur la Fig. \ref{fig6} les oscillations de longue période puis
expliquent que~:

\begin{quote}

\og{}Si la prépondérance de la vitesse est trop considérable, la
période et le point M décrit des cycles peu convergents (Fig. \ref{fig6}),
cycles étudiés pour la première fois par M. Léauté.\fg{}

\end{quote}

Ainsi, quelques années après Léauté, Barbillion et
Cayère ont mis en application le concept de cycle limite de Poincaré
pour résoudre un problème de régulation. Malheureusement, si la
correspondance entre solution périodique et l'existence de courbes
fermées a bien été mise en évidence par chacun de ces
savants aucun d'eux n'a fait le lien avec les travaux de Poincaré et n'a
par conséquent réalisé que les courbes fermées
étudiées étaient des cycles limites.

\section{Les conférences \og{}oubliées\fg{} d'Henri Poincaré sur la T.S.F.}

Jusqu'à présent l'historiographie considérait d'une part que
Poincaré n'avait jamais mis en application le concept de \textit{cycle limite} et attribuait
d'autre part au mathématicien russe Aleksandr' Andronov (1901-1952) le
mérite d'avoir établi une correspondance entre la solution
périodique d'un système auto-entretenu et le concept de cycle limite
de Poincaré dans une note présentée aux \textit{Comptes rendus de l'Académie des sciences} de Paris le 14 octobre
1929. Ainsi, d'après Diner \cite[p. 339]{Diner} :

\begin{quote}

\og{}Au moment même où naît la mécanique quantique, Andronov
participe à l'émergence d'un nouveau paradigme, dont l'acte
fondateur, son travail de diplôme, paraît -- en fran\c{c}ais ! --
dans les \textit{Comptes rendus de l'Académie des sciences} du 14 octobre 1929 : \og{}Les cycles limites de Poincaré et la
théorie des oscillations auto-entretenues\fg{}. Andronov y reconna\^{\i}t,
pour la première fois, que dans un oscillateur de la radiophysique comme
celui de Van der Pol, [{\ldots}] le mouvement dans l'espace des phases est
du type cycle limite, notion introduite par Poincaré en 1880, dans un
contexte purement mathématique.\fg{}

\end{quote}

La découverte d'une série de conférences réalisées par
Henri Poincaré en mai-juin 1908 à l'école Supérieure des
Postes et Télégraphes semble bouleverser quelque peu cette vision
des choses\footnote{Voir Ginoux \textit{et al.} \cite{GinPet} et Ginoux \cite{Ginthesis}.}. En effet, lors
de son dernier exposé intitulé Télégraphie dirigée.
Oscillations entretenues, Poincaré étudie un système entretenu
par un arc chantant. L'arc chantant ou \textit{arc de Duddell} était un dispositif de type
\og{}éclateur\fg{} c'est-à-dire produisant des étincelles qui
engendraient la propagation d'ondes électromagnétiques mises en
évidence par les expériences de Hertz. Tandis que l'éclateur
utilisé par Hertz ne générait que des ondes amorties, l'arc
chantant permettait quant à lui l'établissement d'un régime
stable d'ondes entretenues. C'est pour cette raison qu'il allait \^{e}tre
utilisé au tout début du XX$^{e}$ siècle dans le domaine de la
T.S.F. naissante\footnote{Voir Letellier \textit{et al.} \cite{LetellierGin} et Ginoux et Letellier
\cite{GinLetellier}.}.

Après avoir fournit le diagramme du montage électrique (voir Fig. 7)
dans lequel $x$ représente la charge du condensateur de capacité $1
\mathord{\left/ {\vphantom {1 H}} \right. \kern-\nulldelimiterspace} H$ et
$i$ l'intensité du courant dans le circuit extérieur, Poincaré
établi l'équation différentielle qui caractérise les
oscillations entretenues par l'arc chantant (X).

\begin{figure}[htbp]
\centerline{\includegraphics[width=12.46cm,height=4.68cm]{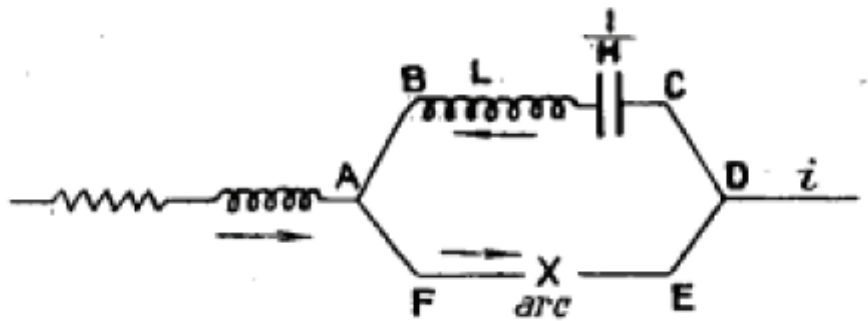}}
\caption[Oscillations entretenues par l'\textit{arc chantant}]{Oscillations entretenues par l'\textit{arc chantant}, d'après Poincaré \cite[p. 390]{P5}.}
\label{fig7}
\end{figure}

Poincaré \cite[p. 390]{P5} explique que ce circuit comprend~\og{}une source de force
électromotrice constante continue $E$, une résistance et une self, et,
en parallèle, d'une part un arc, de l'autre une self et une
capacité.\fg{}. En exprimant alors, au moyen
de la seconde loi\footnote{ Loi des mailles.} de Kirchhoff, la tension dans
la maille ABCDEF, Poincaré établit l'équation différentielle
non linéaire du second ordre des oscillations entretenues par l'arc
chantant

\begin{equation}
\label{eq2}
L{x}''+\rho {x}'+\phi \left( {i+{x}'} \right)+Hx=0
\end{equation}

Il précise alors que \og{}si on suppose connue la fonction $\phi $,
l'équation (Po$_{3})$ donne une relation entre $i$ et ${x}'$ ou entre
$i+{x}'$ et ${x}'$.\fg{}. Poincaré \cite[p. 390]{P5} obtient finalement l'équation suivante qui est tout point analogue à l'équation que le Hollandais Balthazar Van der Pol \cite{Vdp20,Vdp26} établira quelques années plus tard :

\begin{equation}
\label{eq3}
L{x}''+\rho {x}'+\theta \left( {{x}'} \right)+Hx=0
\end{equation}

Bien entendu, cette équation différentielle non linéaire du
second ordre n'est pas intégrable comme le rappelle d'ailleurs
Poincaré \cite[p. 390]{P5} qui établit alors une correspondance entre la solution
périodique, c'est-à-dire, les oscillations entretenues et
l'existence d'une courbe fermée.

\begin{quote}

\og{}On peut construire les courbes qui satisfont à cette équation
différentielle, à condition de conna\^{\i}tre la fonction $\theta $.
Les oscillations entretenues correspondent aux courbes fermées, s'il y
en a. Mais toute courbe fermée ne convient pas, elle doit remplir
certaines conditions de stabilité que nous allons étudier.\fg{}

\end{quote}

Mais, il va plus loin et présente une condition de stabilité des
oscillations entretenues qui est essentiellement basée sur l'existence
d'une courbe fermée~:

\begin{quote}

\og{}\textit{Condition de stabilité}. -- Considérons donc une autre courbe non fermée satisfaisant
à l'équation différentielle, ce sera une sorte de spirale se
rapprochant indéfiniment de la courbe fermée. Si la courbe
fermée représente un régime stable, en décrivant la spirale
dans le sens de la flèche on doit \^{e}tre ramené sur la courbe
fermée, et c'est à cette seule condition que la courbe fermée
représentera un régime stable d'ondes entretenues et donnera lieu
à la solution du problème\footnote{Voir Poincaré \cite[p. 391]{P5}}.\fg{}

\end{quote}

En comparant cette phrase avec la définition d'un cycle limite qu'il
donne dans la \textit{Notice sur les Travaux scientifiques d'Henri Poincaré }faite par lui-même en 1884~:

\begin{quote}

\og{}[{\ldots}] il y a un autre genre de courbes fermées qui jouent un
r\^{o}le capital dans cette théorie~: ce sont les \textit{cycles limites}. J'appelle ainsi les
courbes fermées qui satisfont à notre équation
différentielle et dont les autres courbes définies par la m\^{e}me
équation se rapprochent asymptotiquement sans jamais les atteindre.
Cette seconde notion n'est pas moins importante que la première.
Supposons, en effet, que l'on ait tracé un cycle limite~; il est clair
que le point mobile dont nous parlions plus haut ne pourra jamais le
franchir et qu'il restera toujours à l'intérieur de ce cycle, ou
toujours à l'extérieur.\fg{}

\end{quote}

il apparaît clairement que les courbes fermées dont parle alors
Poincaré sont des cycles limites. Il a ainsi établi, vingt ans avant
Andronov \cite{Andro}, la correspondance entre solution périodique d'un
oscillateur de la radiotechnique et le concept de cycle limite stable qu'il
avait introduit dans ses premiers travaux.

\end{document}